\documentclass[sigconf, screen]{acmart}
 
\AtBeginDocument{%
  \providecommand\BibTeX{{%
    \normalfont B\kern-0.5em{\scshape i\kern-0.25em b}\kern-0.8em\TeX}}}

\copyrightyear{2026}
\acmYear{2026}
\setcopyright{cc}
\setcctype{by}
\acmConference[CHI '26]{Proceedings of the 2026 CHI Conference on Human Factors in Computing Systems}{April 13--17, 2026}{Barcelona, Spain}
\acmBooktitle{Proceedings of the 2026 CHI Conference on Human Factors in Computing Systems (CHI '26), April 13--17, 2026, Barcelona, Spain}
\acmPrice{}
\acmDOI{10.1145/3772318.3791069}
\acmISBN{979-8-4007-2278-3/2026/04}

\usepackage{colortbl}
\usepackage{array}
\usepackage{tabularx}
\usepackage{booktabs}
\usepackage{pdflscape}
\usepackage{booktabs}
\usepackage{graphicx}
\usepackage{lscape}
\usepackage{longtable}
\usepackage{enumitem}

\setcitestyle{maxbibnames=10}

\begin{document}


\title[Results-Actionability Gap]{Results-Actionability Gap: Understanding How Practitioners Evaluate LLM Products in the Wild}



\author{Willem van der Maden}
\affiliation{%
\department{HCI \& Design Section}
  \institution{IT University of Copenhagen}
  \city{Copenhagen}
  \country{Denmark}
}
\email{wiva@itu.dk}

\author{Malak Sadek}
\affiliation{%
  \institution{Centre for Human-Inspired Artificial Intelligence, Cambridge University}
  \city{Cambridge}
  \country{United Kingdom}
}
\email{mfzas2@cam.ac.uk}

\author{Ziang Xiao}
\affiliation{%
  \department{Computer Science}
  \institution{Johns Hopkins University}
  \city{Baltimore}
  \state{Maryland}
  \country{United States}
}
\email{ziang.xiao@jhu.edu}

\author{Aske Mottelson}
\affiliation{%
\department{HCI \& Design Section}
\institution{IT University of Copenhagen}
  \city{Copenhagen}
  \country{Denmark}
}
\email{asmo@itu.dk}

\author{Q. Vera Liao}
\affiliation{%
  \department{Computer Science and Engineering}
  \institution{University of Michigan}
  \city{Ann Arbor}
  \state{Michigan}
  \country{United States}
}
\email{veraliao@umich.edu}

\author{Jichen Zhu}
\affiliation{%
\department{HCI \& Design Section}
  \institution{IT University of Copenhagen}
  \city{Copenhagen}
  \country{Denmark}
}
\email{jicz@itu.dk}

\renewcommand{\shortauthors}{van der Maden et al.}

\begin{abstract}
How do product teams evaluate LLM-powered products? As organizations integrate large language models (LLMs) into digital products, their unpredictable nature makes traditional evaluation approaches inadequate, yet little is known about how practitioners navigate this challenge. Through interviews with nineteen practitioners across diverse sectors, we identify ten evaluation practices spanning informal `vibe checks' to organizational meta-work. Beyond confirming four documented challenges, we introduce a novel fifth we call the \textit{results-actionability gap}, in which practitioners gather evaluation data but cannot translate findings into concrete improvements. Drawing on patterns from successful teams, we contribute strategies to bridge this gap, supporting practitioners' \textit{formalization journey} from ad-hoc interpretive practices (e.g., vibe checks) toward systematic evaluation. Our analysis suggests these interpretive practices are necessary adaptations to LLM characteristics rather than methodological failures. For HCI researchers, this presents a research opportunity to support practitioners in systematizing emerging practices rather than developing new evaluation frameworks.
\end{abstract}






\maketitle

\section{Introduction}
In just a few years, large language models (LLMs) have moved from research labs to production systems, powering everything from marketing copy for local businesses to enterprise software at Fortune 500 companies. This shift has transferred the challenge of evaluation to practitioners who must ensure these systems are effective, reliable, and safe, often without the dedicated infrastructure or methodological guidance that research settings provided. This evaluation gap has emerged as a key bottleneck in production settings~\citep{urlana2024no, gallagher2024assessing, nahar2024beyond, ma2024ismyprompt}, leaving practitioners in a difficult position: they are tasked with building reliable products on a new technological frontier, but are doing so without guiding principles. As these systems mediate crucial aspects of daily life, from healthcare to education, the lack of a standardized foundation to evaluate them creates significant risks that include both costly business missteps and profound harm for individuals and society.

This uncertainty persists despite a wealth of academic inquiry. Recent research flourishes with evaluation frameworks, benchmarks, and novel metrics~\citep[e.g.,][]{liang_holistic_2023, lee_evaluating_2023, elangovan2024considers, pan_human-centered_2024, wang_user-centric_2024, guo_evaluating_2023, wang_evaluating_2024, chang_survey_2024,liu2024ecbdevidencecenteredbenchmarkdesign, an2024leval, liu2024agentbench, wang2024mmlupro, zheng2023judging, zhuo2024bigcodebench, liu2023geval, min2023factscore, kim2024prometheus2}. Yet, it remains an open question how, or even if, this body of work translates into the day-to-day realities of LLM product development. This challenge is not merely theoretical. When Google's NotebookLM team sought to evaluate if AI-generated podcasts, with scripts written by an LLM, were `entertaining,' they discovered that conventional metrics like binary ratings or Likert scales were inadequate~\citep{LatentSpace2024}. Even a single aspect of LLM outputs like humor proved remarkably difficult to assess, with one team member noting that ``humor is contextual... super contextual''~\citep{LatentSpace2024}. This experience raises a crucial question: if well-resourced teams with profound AI expertise struggle with evaluation, what challenges do other, less-resourced, organizations face in assessing their LLM implementations?

While recent studies have documented evaluation challenges and emerging solutions~\citep{zhou-etal-2022-deconstructing, nahar2024beyond, urlana2024no, amershi2019software}, these have focused on either pre-LLM contexts or well-resourced organizations with dedicated infrastructure. Less is known about how practitioners without such support navigate these challenges. Without this understanding, researchers risk developing frameworks that address theoretical problems while missing practical constraints, and practitioners continue reinventing solutions in isolation rather than building on collective knowledge. To unpack this, we investigate:

\begin{itemize}[leftmargin=1.5cm]
\item[\textbf{RQ1}] What are the current evaluation practices for LLM-based products in production settings?
\item[\textbf{RQ2}] What do practitioners describe as their main challenges in evaluating LLM-based products?
\end{itemize}

We conducted semi-structured interviews with 19 practitioners who develop LLM-based products in production settings. These participants predominantly work with foundation models accessed through APIs rather than training their own models (outside of some experiments with finetuning small local models), building applications that must serve real users with specific needs. As such, they evaluate complete systems including user interfaces, retrieval mechanisms, and prompt designs rather than isolated model capabilities. Unlike research labs benchmarking model capabilities through standardized tests, these practitioners must assess context-specific, often hard-to-define qualities while navigating production constraints of limited resources, tight deadlines, and diverse stakeholder demands. Examples include a healthcare documentation system that transcribes clinical conversations (where practitioners must evaluate whether the AI accurately captures meaning without errors that could reverse a diagnosis), educational platforms that generate personalized math stories (where teams assess the balance between student engagement and pedagogical soundness), and enterprise chatbots answering employee questions (where evaluation focuses on maintaining consistency across hundreds of thousands of unpredictable queries).

This paper makes three contributions to HCI and LLM evaluation practice. \textbf{First}, we provide an empirical account of how practitioners evaluate LLM products in production settings. Prior empirical work focused on evaluation of Natural Language Generation (NLG) before the LLM era~\citep{zhou-etal-2022-deconstructing} or documented practices within a single large organization~\citep{nahar2024beyond}; we extend this by studying practitioners across diverse organizational contexts who lack dedicated evaluation infrastructure. While prior work characterizes the widespread reliance on manual testing as a transitional phase awaiting better metrics, we argue that interpretive practices (such as ``vibe checks'') are necessary adaptations to the probabilistic nature of LLMs that must be supported rather than replaced. \textbf{Second}, we identify and conceptualize the \textit{results-actionability gap}: a novel challenge where practitioners successfully collect evaluation data but cannot translate findings into system improvements because they cannot isolate whether failures stem from prompts, retrieval mechanisms, or the model itself. \textbf{Third}, drawing on patterns from successful teams in our study, we contribute actionable strategies to bridge the aforementioned gap through organizational adaptations rather than new metrics. These strategies can support what we observe as practitioners' ongoing \textit{formalization journey} from ad-hoc vibe checks toward systematic evaluation.

For HCI researchers, our findings point to opportunities beyond developing better metrics: supporting the organizational and sociotechnical layers of systematization that practitioners actually need. For practitioners, we provide both validation and direction: validation that current struggles reflect fundamental LLM characteristics rather than poor practice, and concrete direction through three strategies for implementable practices drawn from successful teams and the state-of-the-art. These require no new frameworks or tools, only organizational and process adaptations that our participants discovered through trial and error.
\raggedbottom

\section{Background and Related Work}
We establish key terminology, examine why current evaluation approaches fail in production, and synthesize empirical studies of practice, revealing the need to understand evaluation as an unfolding process rather than a set of barriers.

\subsection{Clarifying Terminology}
\label{sec:clarify}
\subsubsection{What do we mean by `evaluation'?} Our research takes a sociotechnical position and defines LLM evaluation as the process through which teams assess LLM-based systems' fitness for their products' intended goals---a task of determining whether technology satisfies human needs in deployment contexts~\citep{liao_rethinking_2023, weidinger_sociotechnical_2023}. This spans intrinsic approaches (evaluating outputs directly) and extrinsic approaches (measuring effects on task performance)~\citep{gehrmann_repairing_2023}, from early formative assessments through post-deployment monitoring. We further distinguish between model-level evaluation and product-level evaluation. Model-level evaluation focuses mainly on LLM capabilities through benchmarks~\citep[e.g.,][]{chang_survey_2024,liang_holistic_2023,srivastava_beyond_2023}, while product-level evaluation encompasses the complete system including model, user interface, and UX workflows, within the context of use~\citep{raji2020closing,doshiveloz2017rigorous}. Our research focuses on product-level evaluation. In this light, \textbf{evaluation frameworks}, while a somewhat elusive term, refer to structured approaches that prescribe systematic methodologies for conducting evaluation, specifying what constructs to assess, which measurements to use, and how to interpret results---for instance as described by~\citet{tam2024framework}.

\subsubsection{Defining `constructs,' `measurements,' `metrics,' and `criteria'}
Evaluation discussions often conflate what (system goal or aspect) to evaluate, how to measure it, and what constitutes acceptable performance---yet these represent distinct decisions that shape evaluation outcomes in different ways. This distinction is crucial for our analysis: when practitioners report that evaluation ``doesn't work'' or is ``bordering on useless,'' are they struggling to define what matters (constructs), lacking methods to measure it (measurements/metrics), or missing standards for interpretation (criteria)? Understanding where breakdowns occur, and recognizing that solutions at one level may not address problems at another, sharpens our analysis of how evaluation unfolds in practice and what support practitioners actually need. 

\textbf{Constructs} define what goals or aspects of LLM-based systems warrant assessment. Following measurement modeling terminology, these are abstractions describing phenomena of theoretical interest~\citep{allen2001measurement, jacobs2021measurement}---from technical properties (e.g., retrieval accuracy) to experiential qualities (e.g., usefulness in emergency contexts) to systemic outcomes (e.g., fairness across demographics). 

\textbf{Measurements} operationalize these constructs, transforming abstractions into observable data through specific instruments, for instance, automated scoring algorithms, behavioral data, user ratings, or expert assessments~\citep{messick1995, allen2001measurement}. This operationalization necessarily involves assumptions that can introduce mismatches between what we intend to measure and what we actually measure~\citep{jacobs2021measurement}. For instance, `helpfulness' might be operationalized through task completion rates, satisfaction scores, or quality ratings, each capturing different facets while potentially missing others.

\textbf{Metrics} are specific quantifiable measurements that produce numerical outputs. Following measurement theory, metrics are functions that transform system properties into numbers, while measurements are the broader category that includes both quantitative and qualitative assessments~\citep{xiao2023evaluating}. For instance, $F_1$-score is a metric (a mathematical function), while expert judgment is a measurement but not a metric.

\textbf{Criteria} establish performance standards or thresholds, determining what counts as `good' on a given construct. While accuracy is a construct and $F_1$-score is one measurement, requiring ``$F_1 > 0.95$'' establishes a criterion. Setting these thresholds embeds values about acceptable performance levels, a distinct evaluative decision beyond choosing what and how to measure.

\subsection{LLM Evaluation Approaches and Why They Fall Short Today}
\label{sec:evals}
Current approaches to LLM evaluation, from benchmarks to human assessment~\citep{chang_survey_2024}, fail to capture the complexity of production deployments~\citep{mcintosh2025inadequacies}. Benchmarks, such as MMLU~\citep{hendrycks_measuring_2021}, GLUE~\citep{wang2018glue, wang2019superglue}, or aggregates such as HELM~\citep{liang_holistic_2023} evaluate models on well-defined, but often narrowly scoped (e.g., passing an academic test), tasks with clear correct answers, yet LLMs deployed in the real world handle open-ended interactions where success depends on contextual appropriateness. Human evaluation attempts to capture these nuances by having humans rate model outputs using methods such as Likert-scale ratings~\cite[e.g.,][]{Sottana2023Metrics, Zhang2024NewsSumm, Liu2024InstruSum} and pairwise comparisons~\cite[e.g.,][]{zheng_judging_2023, liu2024aligning, bavaresco2024llms}, allowing for the evaluation of multi-faceted, or more subjective qualities such as relevance and coherence. However, these approaches currently lack standardization, and human evaluators can struggle with reliability issues such as disagreement on the interpretation of given constructs~\citep{clark_all_2021} and criteria drift~\citep{shankar_who_2024}. Aside from these challenges, human evaluation is costly and difficult to scale up~\citep{belz-reiter-2006-comparing}. Emerging automated paradigms like LLM-as-Judge aim to scale such subjective evaluations but may introduce systematic biases and recursive validation problems~\citep{thakur-etal-2025-judging, zheng_judging_2023}. Most critically, these approaches are often used to evaluate models in isolation rather than as embedded components of products with specific interfaces, workflows, and user contexts, failing to address the realities of LLM deployment in production systems.

This disconnect exemplifies what has been identified as a ``socio-technical gap''---the persistent divide between what we can measure technically and what matters for actual use \citep{weidinger_sociotechnical_2023}. Researchers have suggested looking to HCI for lessons on bridging this divide, as the field has long grappled with translating technical capabilities into user value \citep{liao_rethinking_2023}. Indeed, recent frameworks have been developed that follow HCI's shift from system-centered to human-centered evaluation~\citep{harrison2007three} by focusing on interaction patterns, stakeholder perspectives, and contextual use rather than isolated outputs and predetermined metrics~\citep{lee_evaluating_2023, collins_evaluating_2024, elangovan2024considers, ibrahim_beyond_2024}. However, whether these theoretically-grounded approaches translate to practice remains unclear. The persistent research-practice divide~\citep{Colusso2017} suggests that even well-designed frameworks struggle against the organizational and resource constraints of production settings~\citep{holstein2019improving}. This gap between proposed solutions and practical adoption motivates our empirical investigation of how evaluation actually unfolds when theoretical frameworks meet production pressures.

\subsection{Empirical Studies of LLM Evaluation in Practice}
\label{sec:empiricalwork}
Empirical work has begun documenting how practitioners approach evaluation in production settings, revealing gaps between academic frameworks and practical realities. Through analysis of public discussions and published literature, researchers have identified evaluation as a critical bottleneck, with practitioners reporting that standardized benchmarks prove ``useless'' for their specific contexts and that human evaluation fails to scale~\citep{mailach2025themes, urlana2024no}. Foundational work on ML engineering established properties that complicate evaluation: component entanglement makes it difficult to trace failures to specific parts of a system, and non-monotonic error propagation means improvements in one area can cause regressions elsewhere~\citep{amershi2019software}. We now examine how these challenges manifest in subsequent work on NLG and LLM evaluation.

\citet{zhou-etal-2022-deconstructing} examined NLG evaluation practices through interviews with 18 practitioners and surveys with 61 participants. Their work revealed that practitioners conflate constructs with the metrics meant to measure them, discussing perplexity when asked about quality goals, for instance. They documented a ``kitchen sink'' approach where 54\% believe in using as many metrics as possible despite recognizing their limitations, with practitioners in an ``in-between phase of just using everything they can find.'' Despite this metric proliferation, 71\% trust manual inspection over automated approaches. They also found that 77\% conduct evaluations primarily to report results in academic papers rather than to guide deployment decisions. Their sample was primarily academic, which partly explains this finding, though Zhou et al. themselves argue this creates pressure toward familiar metrics at the expense of context-specific approaches. This work provides valuable groundwork in surfacing conceptual confusions in evaluation practice, but predates the 2022 LLM boom and focuses on problems rather than solutions, raising the question of whether these patterns persist when practitioners build products rather than publish papers.

More recently, \citet{nahar2024beyond} examined how teams at Microsoft attempt to overcome evaluation challenges when integrating LLMs into software products. Through 26 interviews and 332 survey responses, they documented 19 emerging solutions, from combining qualitative and quantitative metrics to using LLMs as judges. Their findings show evaluation practices in flux: teams spend 76.6\% of effort on manual testing, only 36.3\% have proper evaluation mechanisms, and 46.5\% report ad-hoc metric selection. They also found that many teams have ``no effective methods for evaluating non-deterministic models beyond basic health checks.'' This work provides valuable documentation of both challenges and emerging solutions in the LLM era. However, it focuses on a single company from a software engineering perspective, and catalogs what solutions teams attempt rather than how evaluation unfolds across the product development lifecycle.

Empirical investigations into technical barriers have documented additional challenges. Teams report that the same test yields different results due to non-determinism, and that creating benchmarks requires prohibitively expensive manual labeling~\citep{parnin2025copilot, wang2024quality}. Even standardized benchmarks face what \citet{biderman2024lessons} term the ``Key Problem:'' there is no automatic way to determine semantic equivalence between model outputs. Our work examines how such technical constraints shape evaluation in practice, for instance, how non-determinism complicates stakeholder agreement on what counts as success.

Collectively, this literature establishes that LLM evaluation in production settings is both essential and broken. While previous work has surfaced conceptual confusions in NLG research labs~\citep{zhou-etal-2022-deconstructing} and cataloged challenges and solutions within Microsoft~\citep{nahar2024beyond}, these samples represent well-resourced contexts with established infrastructure. Less is known about practitioners in the practical middle ground: startup teams, self-employed developers, and those simply tasked with making LLM products work without dedicated evaluation support. Our study investigates how evaluation unfolds for these practitioners, examining informal practices, failed attempts, and workarounds that reveal how teams make evaluation work and how these practices shape one another.

\section{Method: Practitioner Interviews}
\subsection{Participants}
We conducted semi-structured interviews with 19 practitioners recruited through professional networks, social media platforms (LinkedIn, BlueSky), and industry conferences. Our sample comprised 5 female and 14 male participants (26\% female), which match typical gender distributions in AI/ML production environments~\citep[e.g.,][]{Deloitte2021WomenAI,UNESCO2021WomenDigital,WEF2025GenderParity,sadek2024interview}. Participants included research professionals ($N=6$), designers ($N=4$), software engineers ($N=4$), data scientists ($N=3$), and marketing professionals ($N=2$). They worked across diverse sectors including healthcare, legal services, education technology, enterprise software, and consumer applications, with organizations ranging from startups to Fortune 500 companies. Participants worked on a range of core LLM functionalities, including retrieval and question-answering ($N=6$), dialogue systems ($N=4$), summarization ($N=3$), content generation ($N=2$), classification ($N=1$), and LLM evaluation or governance ($N=3$). This was a deliberate study design choice to capture a diverse set of perspectives and experiences across sectors. See \autoref{tab:participants} for an overview.

Our study specifically investigates professionals who work with existing foundation models through prompt engineering and product integration, rather than those conducting fundamental model research or modifying architectures. These practitioners face evaluation challenges distinct from traditional benchmarking, as they must assess context-specific, often hard-to-define aspects of LLM performance while navigating real-world constraints of product development. 

\begin{table*}[ht]
\begin{tabular}{llllll}
\hline
\textbf{P\#} & \textbf{Gender} & \textbf{Org. Size} & \textbf{Background} & \textbf{LLM Product} & \textbf{LLM Task} \\
\hline
P1  & Male   & 50-500 & Data Scientist      & Information Retrieval      & Retrieval/QA \\
P2  & Female & 5-50   & Data Scientist      & Content Moderation         & Classification \\
P3  & Male   & >500   & Marketing           & Retail AI Applications     & Dialogue \\
P4  & Male   & 50-500 & Software Engineer   & Enterprise RAG Platform    & Retrieval/QA \\
P5  & Male   & 50-500 & Software Engineer   & Internal AI Assistant      & Summarization \\
P6  & Male   & 50-500 & Software Engineer   & Healthcare Speech System   & Summarization \\
P7  & Male   & >500   & Research            & Automotive Voice Interface & Dialogue \\
P8  & Male   & >500   & Data Scientist      & Marketing Segmentation     & Evaluation \\
P9  & Male   & 5-50   & Research            & Educational Platform       & Generation \\
P10 & Male   & 5-50   & Software Engineer   & Process Automation         & Retrieval/QA \\
P11 & Female & 50-500 & Marketing           & Document Intelligence      & Retrieval/QA \\
P12 & Female & >500   & Research            & Media Content Filtering    & Evaluation \\
P13 & Male   & >500   & Design              & Vehicle Voice Systems      & Dialogue \\
P14 & Male   & <5     & Research            & Training Support Chatbot   & Dialogue \\
P15 & Female & >500   & Research            & Case Management Software   & Governance \\
P16 & Male   & >500   & Research            & Legal Document Analysis    & Retrieval/QA \\
P17 & Male   & 50-500 & Data Scientist      & Law Enforcement Tools      & Summarization \\
P18 & Male   & 5-50   & Design              & Municipal Services         & Retrieval/QA \\
P19 & Female & <5     & Design              & Creative Writing Assistant & Generation \\
\hline
\end{tabular}
\vspace{2mm}
\caption{Participant Demographics ($N=19$). Gender distribution: 14 male (74\%), 5 female (26\%).}
\label{tab:participants}
\end{table*}

\subsection{Procedure}
Data collection took place between February and May 2025. We conducted a pilot study with two ML researchers to refine question clarity and flow in the interview protocol. These pilot interviews were not included in our data analysis. All interviews were conducted remotely via Zoom by the first author, lasted between 45--60 minutes, and were audio-recorded with consent. Participants were not compensated for their time other than early access to the study outcomes. Interviews were conducted by the first author who explored practitioners' full evaluation experience: the methods they use, the challenges they face, and the organizational context in which this work happens. 

Each interview followed a semi-structured protocol exploring participants' experiences with LLM evaluation, covering: (1) current evaluation practices and methodologies that they are using, (2) challenges and constraints in evaluation they face regularly, (3) evolution of evaluation strategies over time, (4) organizational factors affecting evaluation, and (5) tools and resources they used for evaluation at their workplace. 

Data collection continued until patterns became repetitive across interviews and additional participants yielded diminishing new insights~\citep{braun2022thematic}. All interviews were transcribed verbatim for analysis using the paid transcription services of Amberscript\footnote{\url{https://www.amberscript.com/en/}}. Field notes were taken during and after each interview to capture contextual observations and initial analytic insights. All procedures were approved by the University Ethics Review Board.

\subsection{Analysis}
\label{sec:analysis}
We followed the reflexive thematic analysis approach by~\citet{braun2022thematic}. The semi-structured format allowed for flexible exploration of emerging themes and follow-up questions based on participants' responses. The first and second authors initially coded three interviews independently. They separately developed codes inductively. They then met to systematically compare their codes, discuss differences, and collaboratively develop a refined, shared codebook.

Subsequently, all authors discussed this refined codebook and our analytic process to further strengthen the clarity of code definitions. Through this iterative process, codes were refined, consolidated, or added as patterns emerged. For instance, early codes like `organizational tradeoffs' and `organizational tension' were consolidated when we recognized they captured overlapping phenomena around competing priorities. Other codes such as `prompt engineering' and `actionability of results' were added in later iterations as their prevalence across interviews became apparent. Further, discrepancies in how coders interpreted `context-specificity' versus `domain-specificity' were resolved through explicit discussion. The codebook, which went through four iterations during analysis, is available upon request.

Afterwards, the first and second authors divided the interviews and continued to code them based on the updated codebook. Further discussion meetings with all the authors took place to discuss areas of confusion and inconsistency. The first author then derived thematic clusters from the coded transcripts to build up the wider findings and discussion.

\subsubsection{Positionality}
\label{app:positionality}
The research team consists of scholars with backgrounds in human-computer interaction, design research, artificial intelligence systems, and cognitive psychology. Our collective experience includes both academic research and industry collaborations in technology evaluation and development. As researchers trained in interdisciplinary approaches, we are inclined to view technology development as inherently sociotechnical rather than purely technical. Several team members have direct experience working with or studying AI systems in production contexts, while others bring expertise in qualitative research methods and organizational studies. We acknowledge that these backgrounds and our positions within academic institutions shape our approach to understanding practitioners' experiences. 

\section{Findings}
In this section, we first examine why practitioners evaluate LLM products, revealing purposes that span LLM product refinement to organizational politics. Next, we present our findings of RQ1 in the form of ten main evaluation activities, spanning how practitioners execute evaluations [A1-A4], design their approaches [A5-A7], and navigate organizational meta-work [A8-A10]. Our findings of RQ2 reveal five key challenges practitioners face. They cover different aspects related to defining evaluation scope [C1-C4] to implementation barriers and the critical ``results-actionability gap'' [C5].

\subsection{Why Practitioners Evaluate}
Our participants described evaluation as serving multiple, often overlapping purposes. Teams evaluate to understand the capabilities of the specific foundational models their products are built on ---e.g., what models ``can do, what [they] cannot do'' (P2). And they use these insights to guide technical decisions, make business cases, and manage risks. At the technical level, evaluation enables iterative refinement: adjusting prompts based on feedback (P5, P14), comparing whether ``magic happens''---i.e., big jumps in capability improvement---when switching between models (P1), and informing ``trade offs of different training [approaches] or changes in the model architecture'' (P12). At the organizational level, the same evaluations play a role in internal discussions, for instance creating ``data backed cases'' for continued funding (P13) and meeting compliance requirements like the EU AI Act (P15). Participants note that what makes LLM evaluation particularly urgent is the technology's fundamental unpredictability: models are ``pretty unpredictable and they don't follow rules'' (P14) and prone to hallucinations where ``one word wrong can negate the rest of the entire clinical report'' (P6). This unpredictability means evaluation is used as both quality assurance---ensuring ``the right file has been referenced'' (P10)---and risk management, preventing systems from ``blow[ing] in your face'' when pushed beyond simple cases (P7). As P4 summarized, ``the right evaluation strategy has to be proportional to what is at stake''---a principle that shapes how practitioners navigate the complex landscape of practices we now describe.

\subsection{What are the Current Evaluation Practices for LLM-based Products in Production Settings? (RQ1)}
Ten evaluation activities emerged from our data. Below, we organize them into three categories, ordered from actual evaluation to its methodological design and organizational meta-work. They include: 1) how the participants evaluate LLMs  [A1-A4], 2) how they design these evaluations [A5-A7], 3) what organizational meta-work the participants engage around evaluation [A8-A10]. See \autoref{tab:activities} for an overview. 

\begin{samepage}
\begin{table*}[!htb]
\centering
\renewcommand{\arraystretch}{1.0}

\resizebox{\textwidth}{!}{%

\begin{tabularx}{\textwidth}{
    >{\normalsize}p{0.3cm}      
    >{\normalsize\raggedright}p{3.5cm}      
    >{\normalsize\raggedright}p{4.4cm}      
    >{\normalsize\raggedright\arraybackslash}X   
    >{\normalsize}c  
}

\toprule
{\Large\textbf{\#}} &
{\Large\textbf{Activity}} &
{\Large\textbf{Description}} &
{\Large\textbf{Examples}} &
{\Large\textbf{Participants}} \\
\midrule
\multicolumn{5}{l}{\large{\textbf{Evaluation Execution Activities}}} \\
\midrule

\textbf{A1} & Formative \& exploratory checks (``vibe checks'') &
Essential, yet informal first-line evaluations where practitioners conduct unstructured tests of system capabilities before formal metrics &
{\raggedright\textbullet\ ``Hammering'' the system to see errors\newline
\textbullet\ Using imperfect prototypes to expose issues\newline
\textbullet\ ``Dissecting my vibing'' to find quality markers\par}
& $N=12$ \\\addlinespace[6pt]

\textbf{A2} & User evaluation \& feedback &
Continuous collection of end-user feedback across early development and post-launch monitoring &
{\raggedright\textbullet\ In-app thumbs up/down\newline
\textbullet\ Think-aloud protocols \& exit interviews\newline
\textbullet\ Users screenshotting frustrating moments\par}
& $N=17$ \\\addlinespace[6pt]

\textbf{A3} & Expert evaluation as \mbox{continuous} collaboration &
Ongoing engagement with domain experts focused on qualitative feedback and co-design of evaluation criteria &
{\raggedright\textbullet\ SME-led projects\newline
\textbullet\ Red-teaming sessions\newline
\textbullet\ Lawyers checking legal accuracy\par}
& $N=15$ \\\addlinespace[6pt]

\textbf{A4} & Automated tests for integrated systems &
Attempts to apply traditional ML metrics to production systems; practitioners report these are often ``bordering on useless'' &
{\raggedright\textbullet\ Checking HF leaderboards\newline
\textbullet\ BLEU/ROUGE scores\newline
\textbullet\ LLM-as-judge grounding checks\par}
& $N=13$ \\

\midrule
\multicolumn{5}{l}{\large{\textbf{Evaluation Design Activities}}} \\
\midrule

\textbf{A5} & Extracting evaluation criteria &
Thematic refinement that transforms broad concepts into technical constructs &
{\raggedright\textbullet\ Defining ``appropriate to persona''\newline
\textbullet\ UX Questionnaire\newline
\textbullet\ Focusing on 3--4 qualities\par}
& $N=14$ \\\addlinespace[6pt]

\textbf{A6} & Pragmatic metric selection &
Metrics selected for actionability, communication, and practicality &
{\raggedright\textbullet\ Using $F_1$ for stakeholder clarity\newline
\textbullet\ Leaderboard guidance\newline
\textbullet\ Pre-LLM security approaches\par}
& $N=14$ \\\addlinespace[6pt]

\textbf{A7} & Systematizing ad-hoc toolkits &
Shift from informal testing to reusable frameworks (still ongoing for most teams) &
{\raggedright\textbullet\ STPA matrices\newline
\textbullet\ Internal test platforms\newline
\textbullet\ Automated test pipelines\par}
& $N=11$ \\

\midrule
\multicolumn{5}{l}{\large{\textbf{`Meta' Activities Involved in Evaluation}}} \\
\midrule

\textbf{A8} & Alignment activities &
Workshops and sessions for shared understanding and acceptable evaluation approaches &
{\raggedright\textbullet\ Disambiguation sessions\newline
\textbullet\ Governance/ethics workshops\newline
\textbullet\ Negotiating defensible positions\par}
& $N=13$ \\\addlinespace[6pt]

\textbf{A9} & Documenting and sharing practices &
Creating organizational memory through documented practices and reusable assets &
{\raggedright\textbullet\ Metric cards\newline
\textbullet\ Prompt libraries\newline
\textbullet\ Ground-truth datasets\par}
& $N=6$ \\\addlinespace[6pt]

\textbf{A10} & Advocating for evaluation &
Strategic communication and championing needed to prioritize evaluation work &
{\raggedright\textbullet\ Evaluation cases in internal marketplaces\newline
\textbullet\ Acting as ``convinced scientist''\newline
\textbullet\ Securing early involvement\par}
& $N=10$ \\

\bottomrule
\end{tabularx}%
}

\caption{Ten evaluation practices for LLM-based products in production settings}
\label{tab:activities}
\end{table*}
\end{samepage}
\pagebreak 
\vspace*{\fill}
\pagebreak
\subsubsection{Evaluation Execution Activities}
\label{fin:a1}
Our participants employ four main activities when evaluating LLMs, often combining multiple approaches as projects evolve. These span from initial `vibe checks' that provide rapid, intuitive assessment [A1], to continuous user feedback collection [A2], ongoing collaboration with domain experts [A3], and attempts at automated testing from existing paradigms [A4]. 

\bigskip
\noindent\textbf{Vibe checks [A1].}
Twelve participants described beginning their evaluation with informal ``vibe checks.'' These are formative and exploratory assessments that serve as an essential first line of evaluation. As one participant explained, such checks are ``irreplaceable, they are the first line of evaluation'' (P4), a sentiment echoed by another who noted that the ``most important [evaluation] is just individual designer vibing'' (P9).

These initial evaluations are intuitive and variable, with participants struggling to articulate their activities exactly. Some described it as assessing a ``gut feeling'' (P6, P17), while others called it ``prompt vibing'' (P9) or said to be relying on ``entirely subjective and entirely qualitative'' judgments (P14). Given the tacit nature of vibe checks, only three participants were able to articulate what they actually do somewhat clearly. P8 explained how they intentionally used a less-than-perfect LLM product prototype because they ``wanted [the team] to see these errors and immediately think about what types of mitigation they would need to build?'' Similarly, P9 described systematizing their vibe checks by identifying specific quality markers from their intuitive reactions (what they liked or disliked about outputs) and converting these into explicit evaluation criteria to assess generated content.
An interesting exception is P19 who worked on an LLM-based creative writing assistant: \begin{quote}
    {\em ``At the end of the day, we could say that it's based on vibes, but the way of getting to the vibes is very structured and very logically oriented in a spreadsheet. So I had a gigantic spreadsheet where I basically was running the same kind of a prompt [for different characters]. I would see: does that feel right? If I say this out loud, does it sound like a real person speaking? Does it sound like this character that I can hear in my head? [...] I would like test each of those and then like give them a mark. I think is that mark out of ten and then kind of see at the end like [which prompt] is doing best.''}
\end{quote}
While this systematic scoring approach might seem to contradict the ``informal'' nature of vibe checks, P19's framing (``at the end of the day, we could say that it's based on vibes'') reveals how more structured approaches to evaluating LLM products ultimately rely on subjective judgment rather than objective measurement.

This exploratory testing can serve multiple purposes: some participants use it as an ``entry point to getting started and a guide in regards to where should we go when when the next model comes out'' (P6) In other cases, vibe checks serve to identify what aspects should be formalized in structured evaluation, as P17 explains:
\begin{quote}
    {\em ``It's more like to sort of scope what I should be looking for in the responses when I do the initial vibe checking with a model. I would do some manual prompting to get a feel for how we might generate relevant output for the specific use case. Then I will go back and generate even more samples, but in a more structured manner with different settings and different models [...] where I just specify a bunch of different prompts, some different system prompts, some different parameters for temperature and top\footnote{The participant here refers to `Temperature' which adjusts how much randomness is applied when sampling the next token (higher values yield more diverse continuations) and `top-p' (nucleus sampling) limits choices to the most probable tokens whose cumulative likelihood reaches \textit{p}, shaping the balance between creativity and determinism.} and whatnot. And then that will output an Excel sheet with a bunch of different responses [...] that will be what we finally asked the users to rate.''}
\end{quote}

\bigskip
\noindent\textbf{User feedback [A2].}
Seventeen participants described collecting user feedback throughout development, from end-user ``sanity checks'' (P4) to post-launch monitoring. Common mechanisms include in-app ``thumbs up, thumbs down'' (P15, P18) and star ratings (P6), though many question the utility of these simple signals (P4, P5, P9, P14, P15, P16). P16 explains that ``binary feedback is not particularly useful [for legal research] where there are very specific ways that it can go wrong'' and ``simple signals cannot reveal which failure occurred.'' To address this, some teams supplement with dropdown menus specifying ``most anticipated reasons for being dissatisfied'' (P15) or richer qualitative methods like user observation. While prior work confirms practitioners value user feedback~\citep{zhou-etal-2022-deconstructing}, we find these qualitative approaches reveal something binary signals cannot: misalignments between developer and user expectations. P7 recounted:
\begin{quote}
{\em ``We'll have people think out loud... The LLM will just strangely do something roughly related to what the person did. And all of the technical people will be like, oh yeah, this is a bug. And the person is like, oh, cool, that was so cool. It just did what I wanted.''}
\end{quote}

\bigskip
\noindent\textbf{Expert evaluation [A3]} in LLM development is characterized by continuous, collaborative engagement throughout the development lifecycle. Where static benchmarks prove inadequate, teams turn domain experts into a sort of \textit{living benchmarks}---human reference points they repeatedly consult to gauge whether the system is improving or going awry. These experts range from internal specialists like lawyers evaluating legal accuracy (P16) to external consultants assessing creative outputs (P19), from ``embedded travelers'' who help define evaluation approaches (P7, P12) to ``red teamers'' who probe for cases where the system ``is simply not capable of giving a reasonable answer'' (P4). This engagement often begins with ``gut feeling'' checks to assess project viability (P6) and evolves alongside the product. The structure varied. Some described informal sessions where ``teams get together with 2 or 3 people and sit there and work at it together'' (P14). Others ran multi-stage activities that formalized over time, as P3 described:
\begin{quote}    
    {\em ``It started with interviews before the workshops... We presented [our LLM product] back to them and got their feedback. Then my intern made a really nice evaluation platform, and the interior designers went in and would score the accuracy of each one.''}
\end{quote}

\bigskip
\noindent\textbf{Automated evaluation [A4]}, the last main evaluation activity, also takes several forms in practice, though none has achieved the reliability practitioners seek. For initial model selection, some teams (P2, P6, P7, P10, P11, P17) consult existing benchmarks and leaderboards, using Hugging Face to ``narrow down my search'' (P1). However, none reported running comprehensive benchmark suites like MMLU or SuperGLUE for production evaluation---participants found such benchmarks ``bordering on useless'' (P10) for their specific contexts.  Teams also attempt to apply traditional ML metrics such as $F_1$-scores (P2, P15), BLEU (P10), or BERT/ROUGE scores (P17) to their systems, though creating appropriate test data remains a burden (a challenge we discuss in C2). More recently, some teams have adopted LLM-as-judge approaches. Three participants (P6, P10, P15) actively used LLMs to evaluate their products---for instance, checking whether chatbot answers were ``grounded in the retrieved information'' (P15) or scoring outputs for ``truthfulness and completeness'' (P6). Others considered but rejected these approaches, with P8 viewing LLM-as-judge as an untraceable ``black box.''

\bigskip
\noindent\textbf{Summary of evaluation types} A1--A4 reveal a notable pattern: while practitioners aspire to automated, scalable evaluation [A4], they currently rely heavily on human judgment---whether through developer/designer intuition [A1], user feedback [A2], or expert assessment [A3]. This dependence on subjective, context-specific evaluation methods sets the stage for understanding how our participants attempt to design their evaluation approach. 

\subsubsection{Evaluation Design Activities}
Beyond execution, practitioners must design their evaluation approaches---deciding what to measure and how. This involves extracting testable constructs from qualitative observations [A5], selecting measurements that balance rigor with practical constraints [A6], and attempting to codify ad-hoc methods into reusable frameworks [A7].

\bigskip
\noindent\textbf{Construct extraction [A5].}
Fourteen participants described determining what to evaluate through thematic refinement of qualitative data. Prior work documents that practitioners conflate quality criteria with metrics~\citep{zhou-etal-2022-deconstructing}; we find teams actively work to disentangle them through construct operationalization. For example, ``appropriate'' becomes ``appropriate tone and persona for context'' (P12), ``useful'' becomes ``saves money or improves quality'' (P4), ``engaging'' becomes ``storyline quality and interaction effectiveness'' (P9). Teams derive these constructs from multiple sources: user feedback sessions, developer observations, stakeholder discussions, and domain expert input. As P7 noted, when dealing with subjective qualities, ``we need to be on the same page when we evaluate something. What is it that we're evaluating?''

This refinement typically begins with high-level concepts that practitioners acknowledge as ``fluffy'' yet critical (P4, P15). One participant vividly described the journey from intuitive judgments to explicit constructs:
\begin{quote}
    {\em ``We've got this big list of quality markers that I've noticed over time. Like when I like something or don't like something, I'm trying to sort of dissect my vibing---what is it that I don't like, when I think it's crap? [...] And then I give that as evaluation criteria... What I've found to be most useful is where it attempts the evaluation and kind of pulls out specifics, but I'm the one ultimately deciding.'' (P9)}
\end{quote}
P9 elaborated that LLM-based evaluation works best when it ``gives examples of how it's strong or weak in that area'' rather than numerical scores---``what you're really looking for is where was the specific experience that was subpar and what made it subpar.''

\bigskip
\noindent\textbf{Metric selection [A6].}
Once teams know \textit{what} to evaluate, they must decide \textit{how} measure it. Prior work found that faced with metrics they do not fully trust, practitioners adopt a ``kitchen sink'' approach---``you kind of just throw it all at the wall''~\citep{zhou-etal-2022-deconstructing}. Our participants responded differently to this uncertainty: rather than adding more metrics, they selected fewer based on actionability. P1 chose measurements ``that I can actively do something to improve'' and were ``much easier for me to interpret.'' To make these selections, teams blend multiple influences: adapting measurements from prior non-LLM systems (P4), consulting external sources (P9, P10), or inventing ``what would be the most valid criteria'' for their specific context (P6). P2 explained:
\begin{quote}
   {\em ``We chose the $F_1$-score because it's the easiest way to have one number. Sometimes it becomes too complex. We could present recall and precision separately, go through all the individual metrics. But that doesn't help people understand. Our goal was to have one number that can be understood, even though they don't understand what an $F_1$ score is necessarily.'' (P2)}
\end{quote}

\bigskip
\noindent \textbf{Systematizing ad-hoc toolkits [A7].}
Eleven participants described attempts to codify their evaluation approaches into reusable infrastructure, moving from informal ``trial and error'' methods (P4) and individual vibe checks toward structured, repeatable approaches. Most reported they had not yet achieved their desired level of formalization. As P18 explained, they are ``doing it very much on a case by case basis right now'' with the hope that ``at some point we will start to learn some patterns.'' P8 described one successful systematization:
\begin{quote}
    {\em ``I personally really like system theoretic process analysis. You basically identify the things you don't want to happen, then you create a set of hazards. Those constraints become your requirements, which become the specific requirements for the product. [...] Creating tests and a requirement matrix so that every test aligns to a set of requirements. [...] If the test fails, you know which requirement did not function, and you can immediately intervene.''}
\end{quote}
Participants pursued systematization for several reasons: avoiding the inefficiency of ``reinvent[ing] the wheel'' (P15) and escaping the cost of remaining reactive. Without systematic approaches, P8 noted, teams get ``stuck doing somewhat of an assurance job saying, look, folks, you didn't build a good product. Now you have to rebuild it.'' Timing also proved critical. P6 stated: ``you have to have the [evaluation] tooling in place before you start creating the functionality. Then you get the benefit of the tooling all the way.'' However, participants described their current state as ``not at all scientific'' (P14), ``definitely not in a place where it's been formalized'' (P6), with evaluations that had ``not been systematic'' and therefore ``not fed into the design at all'' (P15). We refer to this progression from ad-hoc practices toward systematic evaluation as a \textit{formalization journey}, a concept we return to in the discussion.

\bigskip
\noindent\textbf{Summary of evaluation design practices [A5--A7]} These activities reveal practitioners caught between pragmatic necessity and systematic aspiration. They transform intuitive quality judgments into measurable constructs, select measurements for stakeholder communication over technical rigor, and attempt to codify these discoveries into reusable frameworks---though most remain stuck repeating others' expensive lessons. The gap between where teams are (reactive, ad-hoc) and where they want to be (proactive, systematic) defines the current state of LLM evaluation design.

\subsubsection{The `Meta' Activities That Shape Evaluation}
\label{sec:metawork}
The design and execution of evaluations does not happen in isolation and is shaped by forces at the organizational level. The group who carries out evaluation need to communicate, compete for resources, and coordinate with other teams in the company. This meta-work is not unique to LLM products. However, its existence is entangled with the technical complexity and methodological uncertainty of LLM evaluation manifested in [A1-A7]. Below we only highlight the meta-work directly influencing LLM evaluation. 

\bigskip
\noindent\textbf{Alignment work [A8].}
Thirteen participants described disambiguation sessions and collaborative workshops (formal or informal) where teams surface different perspectives and negotiate shared evaluation constructs:
\begin{quote}
    {\em ``I held this kind of disambiguation session where I said, okay, what are we talking about? [...] I realized we were all talking about completely different things actually. The designers, the product managers, the technical people, they were talking about actual user experience, the AI engineers and me, we're talking about something very different [...] Eventually we just landed on groundedness [... because it] was the one that the product managers could understand the most: is the answer provided grounded in the documents that the customers provided?''} (P15)
\end{quote}
These shared constructs then feed directly into evaluation design [A5, A6]. Without alignment, teams risk measuring different things and ending up with in-actionable results [C5].

\bigskip
\noindent\textbf{Documentation and sharing practices [A9]} are reported by six participants who create artifacts to preserve evaluation methods for future use. The forms varied: P15's team developed ``AI performance metrics cards'' standardizing definitions across products; P11 built ground-truth datasets with ``guidelines for evaluators as to how you would rate''; P13 designed a human feedback system explicitly as ``a blueprint... that could also be transferable across teams''; and P9 maintains ``prompt chains'' in a Google doc, describing the goal as ``operationalizing your own vibe, trying to extract what you feel in a way that can be a criteria that can maybe be updated later.'' P15 articulated the broader rationale: creating ``a portfolio of records made of decisions'' about why certain measurements were chosen ``so that you don't have to reinvent the wheel.'' Yet documentation remains culturally difficult. As P15 noted, ``people are not used to documenting things, and LLM evaluation requires really good documentation.''

\bigskip
\noindent\textbf{Advocating for evaluation [A10]} emerged as essential organizational work. Ten participants described extensively advocating to secure resources and legitimacy for evaluation activities. P15 framed the work as ``95\% communication... people who can communicate well are the ones who win at the end of the day.'' This advocacy shapes not just whether evaluation happens, but when and how. P13 explained how early-stage resource constraints force strategic choices:
\begin{quote}
    {\em ``When you're at an early stage of a project like this, you're really oriented at making the best case possible for the internal ideas marketplace. And that means that user centered [evaluations] are only important so far as they advance the project, so that we can then actually put more capital into the user studies.''}
\end{quote}
Others work to shift organizational timing, trying to be ``in the room when they're conceptualizing'' products rather than ``tacking safety on as a mitigation at the very end'' (P12).

\bigskip
\noindent\textbf{Summary of evaluation meta-work [A8--A10]} These activities reveal that evaluation extends far beyond technical measurement to encompass essential organizational work.  This meta-work is not peripheral ---it is social infrastructure where teams devise evaluations and attempt to improve their products.

\subsection{What do Practitioners Describe as Their Main Challenges in Evaluating LLM-based Products? (RQ2)}
\label{sec:challenges}

\begin{samepage}
\begin{table*}[!htb]
\centering
\renewcommand{\arraystretch}{1.0}
\resizebox{\textwidth}{!}{%
\begin{tabularx}{\textwidth}{
    >{\normalsize}p{0.3cm}      
    >{\normalsize\raggedright\arraybackslash}p{3.5cm}      
    >{\normalsize\raggedright\arraybackslash}p{3.5cm}      
    >{\normalsize\raggedright\arraybackslash}X   
    >{\normalsize}c  
}
\toprule
{\Large\textbf{\#}} &
{\Large\textbf{Challenge}} &
{\Large\textbf{Description}} &
{\Large\textbf{Examples}} &
{\Large\textbf{Participants}} \\
\midrule
\multicolumn{5}{l}{\textbf{\large{Challenges related to defining evaluation objectives \& scope}}} \\
\midrule

\textbf{C1} & Aligning on evaluation objectives & Teams cannot agree on what they are trying to achieve with evaluation &
{\raggedright\textbullet\ ``Talking about completely different things''\newline
\textbullet\ Data scientists vs.\ AI engineers disagreeing\newline
\textbullet\ Settling for ``least rejected'' options\par}
& $N=10$
\\\addlinespace[6pt]

\textbf{C2} & Establishing clear and meaningful constructs & Defining what to measure and what constitutes ``good'' &
{\raggedright\textbullet\ ``I don't know what `good' looks like'' (i.e., what constructs to use)\newline
\textbullet\ Uncertain if ``40\% is good or bad'' (i.e., success criterion to use in this case) \newline
\textbullet\ Limited to ``does it look reasonable?''\par}
& $N=13$
\\\addlinespace[6pt]

\textbf{C3} & Deciding on evaluation approach & Choosing viable methods given constraints and uncertainty &
{\raggedright\textbullet\ Difficulty imagining user questions\newline
\textbullet\ Testing only ``happy path'' scenarios\newline
\textbullet\ Relying on ``vibe, feeling'' over data\par}
& $N=13$
\\

\midrule
\multicolumn{5}{l}{\textbf{\large{Challenges related to implementing evaluations}}} \\
\midrule

\textbf{C4} & Technical and operational barriers & Infrastructure, non-determinism, and human resource constraints &
{\raggedright\textbullet\ Models working ``two times in five''\newline
\textbullet\ ``Humans don't scale''\newline
\textbullet\  Inference not feasible on current hardware\par}
& $N=10$
\\\addlinespace[6pt]

\textbf{C5} & The Results-Actionability gap & Translating evaluation outcomes into improvements &
{\raggedright\textbullet\ Not knowing how to use evaluation data\newline
\textbullet\ ``Collecting data without a plan''\newline
\textbullet\ Easier to ``redo the whole thing''\par}
& $N=17$
\\

\bottomrule
\end{tabularx}%
}
\caption{Five evaluation challenges for LLM-based products in production settings}
\label{tab:challenges}
\end{table*}
\end{samepage}

Throughout the range of evaluation activities (A1--A10), our participants consistently encounter challenges that complicate or undermine their efforts. Many of these challenges have been documented before: aligning stakeholders, establishing constructs, choosing methods, and overcoming technical barriers all appear in prior work~\citep{zhou-etal-2022-deconstructing, nahar2024beyond}. We cover these briefly [C1--C4], confirming they persist from NLG into LLM contexts and extend beyond single organizations like Microsoft. However, our analysis also surfaces a challenge that prior work has not explicitly theorized: the ``results-actionability gap'' [C5], where teams gather evaluation data but cannot translate findings into concrete improvements. This gap affected 17 of our 19 participants and receives extended treatment below, as it helps explain why evaluation struggles persist even when teams overcome the preceding challenges.

\subsubsection{Documented challenges: why designing and executing evaluations remains hard}
\textbf{Aligning on evaluation objectives [C1]} emerged as a major struggle for ten participants. Teams dedicate entire sessions to this work [A8], reflecting how difficult reaching shared understanding proves to be. The challenge stems from stakeholders attending to fundamentally different concerns. P16 described four competing levels:
\begin{quote}
    {\em ``My employer cares most about not having hallucinations... But then the vendor might care about something different and the user probably cares about something different... and then of course there's the technical people who are just making sure that your hyperparameters or whatever and your rank system is good. So there's like four levels of [evaluation]. That's by different people, right?''}
\end{quote}
These groups struggle to translate between their framings: a technical improvement from ``80\% to 95\%'' may be ``a cool technical achievement,'' but ``the relationship between the technical scores and the user, like the UX, is not clear'' (P16). Without shared language, teams settle for compromise---adopting constructs like ``groundedness'' not because it best captures quality, but because it was ``the [construct] they were rejecting the least'' (P15). However, agreeing to measure ``groundedness'' still leaves open what groundedness means---which poses the next challenge.

\textbf{Establishing clear and meaningful constructs [C2]} proved difficult for thirteen participants with P12 calling it ``the toughest part'' of evaluation. 
Prior work documented teams struggling to determine what counts as correct~\citep{nahar2024beyond}. Our analysis points to a specific source of this difficulty: the absence of reference points. As P12 explained, ``I don't know what good looks like before we start kind of release it into the wild.'' This challenge is compounded by context-dependency; constructs that seem straightforward shift meaning across domains:
\begin{quote}
    {\em ``How do you handle a block list when you're thinking of, like, a music catalog? Um, you know, where lyrics have every bad word that exists. [The construct of] `appropriate content' that might work elsewhere breaks down when there's an artist called Child Abuse, for example, which is probably ironic.''}
\end{quote}
Faced with this complexity, teams often retreat: ``they said, oh, it's too hard to measure relevance. It's too hard to measure'' (P15). Without clear constructs, extracting evaluation criteria from qualitative data [A5] becomes guesswork rather than systematic refinement.

\textbf{Identifying viable evaluation approaches [C3]} proved equally difficult ($N=13$). Teams struggle with basic methodological questions: whether to use LLM-as-judge approaches or end-user feedback, how to structure evaluations, whom to involve. Yet there is ``no systematic'' way to choose among methods (P4). When teams look for guidance, they encounter a gap: ``You find all these frameworks that are all academic and none of them have been tested in product. And you're like, I don't know which is best'' (P12). As a result, ``we end up just kind of building our own things'' (P12), but without confidence these approaches are sound. As P5 explained, ``We're finding it kind of difficult to find out exactly how is a good way to do this... We ended up with just some sort of vibe, feeling.'' This reflects what \citet{zhou-etal-2022-deconstructing} described as an ``in-between phase of just using everything they can find.'' The uncertainty has consequences: some teams ``launch without evaluating'' entirely (P15), while others remain stuck, unable to systematize their ad-hoc toolkits [A7] into reusable approaches.

Finally, \textbf{technical and operational barriers [C4]} prevent evaluations from occurring at all ($N=10$). Infrastructure proves inadequate, as P2 noted: ``inference on LLMs is not feasible on any municipality's current hardware.'' Even well-resourced teams encounter instability, with APIs yielding constant errors that derail testing (P10). Human evaluation offers no escape from these constraints: ``humans don't scale'' (P6), specialized domain expertise is expensive to recruit (P1, P5, P18), and internal testers resist evaluation tasks they view as ``actually quite boring and [time consuming]'' (P5). These barriers compound across evaluation activities, limiting user feedback collection [A2], expert collaboration [A3], and automated testing [A4] alike.

\subsubsection{Novel challenge: The Results-Actionability Gap}
The final and perhaps most significant challenge practitioners face is bridging the \textbf{``results-actionability gap'' [C5]}: the difficulty of translating evaluation outcomes into concrete, actionable steps for product improvement ($N=17$). The core of this challenge is a fundamental question that practitioners struggle to answer: ``How do we actually then use the information to inform the decisions going forward?'' (P6).

This actionability gap is caused by two primary factors. First, the evaluation results themselves are often ambiguous. Practitioners find that qualitative feedback, such as a subjective ``vibe... does it feel right?'' (P14) or a context-dependent user preference (P3), does not easily translate into a deterministic plan where ``if outcome X then we take action Z'' (P6). This is because such feedback describes holistic impressions rather than specific components. P14 recounted testing a chatbot with diverse users:
\begin{quote}
    {\em ``We've given it to a Black younger woman. And she said, no, it just seems really patronizing. And we just have to go, oh, okay. Well, it didn't seem patronizing to us, but it does to you. So we need to do something about that.''}
\end{quote}
The feedback is valid---but ``patronizing'' does not indicate whether to adjust the prompt's tone, change the persona instructions, modify response length, or restructure the interaction flow. The team knows there is a problem but not which component to change.
Second, even when a result is specific (such as a low score on a metric) it is often difficult to trace it back to a root cause. LLM-based systems involve many interacting variables---prompt wording, model selection, temperature settings, retrieval parameters, embedding models, context windows---any of which could be responsible for a poor outcome. P8 described this directly:
\begin{quote}
    {\em ``If you're using an embedding model or a ranker or RAG and you're like, oh, it's actually not necessarily the LLM call itself that's problematic. It's the combination of these six things. Then you have to develop a whole separate set of tests until you can finally identify a potential root cause.''}
\end{quote}
P17 framed this as a parameter problem without guidance: ``There are tons of parameters that you can tune... Which dials do I change?'' As P18 summarized: 
\begin{quote}
    {\em``There are all these degrees of freedom that stack up and make the evaluation very unclear. You don't necessarily know what you have to do next. When you have an evaluation scale and you end up with a 4.6 out of ten, if you don't know what caused that, then it's very difficult to iterate on making it better.''}
\end{quote}
The consequences of this actionability gap are significant, as one participant articulated:
\begin{quote}
    {\em ``If we get a groundedness score of 0.5, what do we do? [The developers] don't want to answer. The AI engineer keeps pushing for a plan, but the best I get is ``we'll monitor for trends.'' When I ask if we'd pull the product for bad scores, no one answers. We're collecting data without a plan for what we're going to do with it---and potentially we just ignore it if we don't like it.'' (P15)}
\end{quote}
This confusion makes the entire evaluation process feel fruitless, with another participant noting that it is ``very, very difficult to do it in a way that actually gives valuable results right now'' (P18). At worst, when iterative refinement proves impossible, practitioners find that sometimes it is ``just easier to redo the whole thing'' (P9). But often, neither refinement nor rebuilding happens. P8 described what occurs when evaluation findings arrive late in development: ``You're just pointing out a problem without a solution... [and] everybody just goes: Nope! We're pushing it live and we'll deal with it later.'' The consequence is lasting: ``You're stuck with known issues that nobody's ever going to invest in'' (P8). Evaluation has occurred, problems have been identified, but the results-actionability gap means nothing changes.


\section{Discussion}
\label{sec:disc}
This study investigated how practitioners evaluate LLM-based products in production settings. We interviewed 19 practitioners who develop LLM-based products across diverse sectors. Our analysis identified ten evaluation practices [A1--A10] spanning evaluation execution, evaluation design, and the organizational meta-work that shapes both. We also identified five challenges that complicate these practices. 

Four of these challenges [C1--C4] confirm findings from prior studies~\citep{zhou-etal-2022-deconstructing, nahar2024beyond}, revealing a persistent pattern: across all three studies, practitioners rely on manual testing and interpretive methods rather than metric-based evaluation. Prior work treats this reliance as a problem to solve through better frameworks and training. We interpret this differently: these practices may not be problems to solve, but necessary adaptations to LLM characteristics that warrant support rather than replacement. We observed teams attempting to systematize their practices [A1--A10], progressing from ad-hoc vibe checks toward reusable evaluation approaches. We call this the \textit{formalization journey}. Most teams are somewhere along this journey, navigating it through trial and error without support. HCI has established methods for exactly this kind of work, presenting a research opportunity discussed in \autoref{sec:rec}.

The fifth challenge, the results-actionability gap [C5], has not been previously documented despite being experienced by 17 of 19 participants. Practitioners gather evaluation data but cannot translate findings into concrete improvements. They expect evaluation to work like traditional ML or software testing: a low score points to the component that needs fixing. LLM products are different: multiple components interact in unpredictable ways. A low score does not reveal whether the problem lies in the prompt, the temperature, or the retrieval system.

To understand why the same challenges reemerge across studies and why practitioners consistently turn to interpretive methods like vibe checks, we first situate our findings within existing empirical work on LLM evaluation (\autoref{sec:situating}). We then theorize four structural factors that may be inherent to the LLM paradigm that shape the evaluation challenges (\autoref{sec:factors}). From this analysis, we discuss implications for HCI research and recommendations for practitioners (\autoref{sec:rec}).

\subsection{Situating Our Findings in the Practical LLM Evaluation Literature} \label{sec:situating} 
We build on previously discussed empirical literature, in particular Zhou et al.'s study of NLG evaluation~\citep{zhou-etal-2022-deconstructing} and Nahar et al.'s analysis of LLM evaluation at Microsoft~\citep{nahar2024beyond}. Our participants lack the deep ML expertise of dedicated NLG research groups and the extensive infrastructure available at Microsoft. Despite these constraints, they confirm that fundamental evaluation problems persist from NLG to LLM contexts: practitioners continue to conflate quality criteria with measurements and struggle to articulate what constitutes `good' output, confirming patterns documented in evaluation pre-LLM systems~\citep{zhou-etal-2022-deconstructing}. Similarly, the emerging solutions \citet{nahar2024beyond} identified at Microsoft---such as defining custom metrics through expert collaboration---appeared independently across our sample. By comparison, our analysis extends this prior work by identifying three specific mechanisms practitioners use to navigate these constraints:

\begin{enumerate} 
\item \textbf{Reframing manual testing as an inherent heuristic.} Both \citet{nahar2024beyond} and \citet{zhou-etal-2022-deconstructing} document heavy reliance on manual testing. While these prior works frame this as problematic~\citep{zhou-etal-2022-deconstructing} or burdensome~\citep{nahar2024beyond}, our analysis reveals these ``vibe checks'' serve as essential first-line evaluation. Participants described them as ``irreplaceable'' (P4) assessments that capture qualities that formal metrics miss. The persistent reliance on manual testing across studies suggests this may be inherent to evaluating LLM-based systems rather than a methodological weakness to overcome. We therefore advocate supporting and systematizing these intuitive practices rather than attempting to \textit{replace} them.

\item \textbf{From ``better'' metric selection to actionable construct extraction.} Prior work identified that practitioners conflate quality criteria with measurements~\citep{zhou-etal-2022-deconstructing}, a confusion also present in Microsoft's LLM teams~\citep{nahar2024beyond} and in our sample. Responses vary, from ``kitchen sink'' approaches where teams try every available metric~\citep{zhou-etal-2022-deconstructing}, to structured research phases for defining custom metrics that combine subjective measures with objective ones~\citep{nahar2024beyond}. Our participants take a further step: they accept that metrics often fail for their contexts and use qualitative inquiry not as a complement to measurement, but as the source from which evaluation constructs emerge. These constructs are then evaluated through interpretive methods common in HCI such as reflective practice (P9), think-aloud protocols (P7), and disambiguation workshops (P15) rather than measurement. This may be a necessary adaptation to LLM characteristics rather than faulty practice.

\item \textbf{The overlooked role of organizational meta-work.} Finally, we surface the organizational meta-work that evaluation requires [A8–A10]. Alignment, documentation, and advocacy are not unique to LLMs---they characterize complex technology development generally~\citep{amershi2019software}. However, prior LLM evaluation studies treat these activities as background noise, even though they are entangled with evaluation itself—shaping what gets measured and how. Without alignment, stakeholders within a team may be evaluating different constructs without realizing it. Without documentation, teams risk repeating discoveries others have already made. Without advocacy, evaluation may get deprioritized or sidelined entirely. Therefore, efforts to improve LLM evaluation should account for this organizational layer.
\end{enumerate}

\noindent Taken together, prior work frames practitioners' reliance on manual and qualitative methods as problems to solve: \citet{zhou-etal-2022-deconstructing} describe an ``in-between phase'' awaiting better metrics; \citet{nahar2024beyond} document teams searching for more effective measurement approaches. Improving metrics, criteria, and methods for developing them is valuable work. Additionally, we suggest that interpretive approaches may not be purely transitional; in some contexts they may be necessary adaptations where metrics-based evaluation cannot provide actionable signal. The following section examines four structural factors that we argue explain why this is the case.

\subsection{Incompatibility of Metrics-based Evaluation with LLM Realities: Factors Driving Interpretation}
\label{sec:factors}
From our analysis, we identify four factors that we argue are typical to evaluating LLM products. This is not an exhaustive list, but we theorize these conditions underlie the challenges practitioners described. We suggest they warrant consideration in both future research and evaluation practice. Below, we examine how each complicates the evaluation process:
\begin{enumerate}
    \item[F1] \textbf{The mismatch between general-purpose models and specific contexts.} Unlike traditional workflows where AI/ML models are trained for specific tasks, off-the-shelf LLMs function as general-purpose engines. This creates a structural misalignment: model providers optimize for broad capabilities, while practitioners require reliability in specific contexts. Since practitioners typically operate as integrators via APIs rather than model trainers, they are precluded from tailoring the underlying model to their requirements, forcing them to rely on inference-time adaptations such as system prompting.\footnote{Domain-specific models (e.g., specific checkpoints or specialized providers) are emerging, but participants in our sample rely on ``off-the-shelf'' models rather than having the capacity or resources to train their own.} This complicates evaluation because a general-purpose model may perform adequately but not excellently at specific tasks, and can unexpectedly drift into unrelated capabilities rather than staying within product requirements. As P14 noted, standardized benchmarks capture only ``a small part of the probability distribution,'' leaving teams with no guarantee that a passing score on a general benchmark implies reliability for their specific use case. Consequently, teams are forced to ignore established benchmarks and build bespoke test suites from scratch.

    \item[F2] \textbf{Non-determinism combined with absent ground truth.} This factor creates what participants experienced as a dual epistemological problem: establishing adequate test data and determining ``correctness.'' Unlike traditional ML where teams could derive test sets directly from the same source of training data~\citep{nigenda2022amazonsagemakermodelmonitor, shergadwala2022humancentricmodelmonitoring}, practitioners here must generate test data \textit{out of nothing}. This is complicated by the fact that ``correctness'' is often perspectival rather than objective; different users legitimately disagree on quality, making the search for universal evaluation criteria conceptually problematic~\citep{aroyo2015truth, davani2022dealing}. When outputs work ``two times in five'' (P14), evaluation becomes an exercise in subjective interpretation rather than objective measurement. Consequently, ``success'' ceases to be a static, pre-agreed threshold (e.g., ``$F_1$ must exceed 0.9'') that persists across iterations. Instead, it becomes a negotiated agreement that must be debated and recalibrated within each project context---a dynamic made even more volatile by the rapid pace of model advancement.  

    \item[F3] \textbf{The failure of familiar quantitative approaches.} Traditional software testing relies on clear diagnostic signals, an assumption that practitioners found violated by LLMs. Prior to this era, organizations utilized ML evaluation infrastructure where specific signals (e.g., drift detection alerts) triggered clear actions (e.g., retraining on new data)~\citep{nigenda2022amazonsagemakermodelmonitor}. However, practitioners found that such pipelines now yield numbers without clear next steps for improvement. Because familiar metrics (like accuracy) fail to correlate with user experience in open-ended contexts (e.g., LLM conversations or LLM-supported coding), the established feedback loop breaks down. This forces a methodological pivot: practitioners turn to qualitative methods (vibe checks, expert judgment) out of necessity, yet often dismiss these practices as ``not scientific'' (P14). This concern stems not from the methods themselves being invalid, but because practitioners trained in quantitative paradigms lack the frameworks for comprehensive qualitative analysis that would reveal these approaches as rigorous evaluation methods.

    \item[F4] \textbf{Adoption barriers for emerging frameworks.} The overwhelming rate of LLM deployment has caused a proliferation of new evaluation platforms. However, P12 noted the ``cost of learning'' them often outweighs the perceived benefit. This complicates evaluation by forcing teams to prioritize local, expedient solutions (like manual spreadsheets) that become unmaintainable at scale. The result is ``technical debt''~\citep{cunningham1992}, preventing the formalization of their evaluation practices.
\end{enumerate}
These factors explain why the \textbf{formalization journey} is so difficult: practitioners are not failing to measure; they are operating in a context where measurement often fails to provide actionable signal. By recognizing that these challenges stem from the fixed nature of the technology rather than practitioner incompetence, we can better target future research.

\subsection{Implications for HCI Research and Evaluation Practice}
\label{sec:rec}
We first discuss opportunities for HCI research, then offer strategies for practitioners.
\subsubsection{For HCI: The Formalization Journey as Research Opportunity}
Our participants are at various points along a formalization journey, progressing from ad-hoc vibe checks toward systematic evaluation. The practices they develop along the way (e.g., examining intuitive reactions for quality markers, facilitating sessions to align on criteria, refining approaches through pattern recognition) resemble established HCI methods. Yet both practitioners and prior research treat these interpretive approaches as inferior: our participants dismiss their own work as ``not at all scientific'' (P14), while prior studies frame such methods as transitional, awaiting better metrics~\citep{zhou-etal-2022-deconstructing, nahar2024beyond}. HCI has long established these as appropriate methods for contexts where measurement alone cannot capture what matters. This suggests an opportunity: HCI can help scaffold and systematize these emerging practices---a role the field has played before. 

Prior conversational AI research documents similar patterns of practitioners following ``a blind process where they [type] a bunch of stuff, they train their model and then they hope that it didn't break'' while developing evaluation practices~\citep{sadek2023cui}. When chatbots and voice assistants emerged, HCI helped transform such ad-hoc testing into systematic conversational design~\citep{sadek2023cui, elshan2022, roedl2013}. The factors that make LLM evaluation unique compared to previous AI/ML technologies---the conditions driving practitioners toward interpretation---are what HCI methods in many cases were developed to address. For instance, contextual inquiry assumes technology must be understood in situ rather than through abstract metrics, addressing the gap between general-purpose models and specific contexts [F1]. Participatory design embraces multiple perspectives rather than seeking singular ground truth, working with the subjective nature of LLM quality [F2]. Ethnographic approaches expect qualitative data rather than clean quantitative signals, matching what practitioners encounter [F3]. And HCI's tradition of discount methods provides lightweight approaches when formal frameworks prove too heavy [F4]. These are not workarounds for difficult conditions---they are methods designed for these realities. 

These conditions suggest a reframing of HCI's role in LLM evaluation. Rather than developing new evaluation frameworks that practitioners must learn and will likely abandon, the field could focus on recognizing and systematizing the practices teams already employ. This means creating bridges between practitioner language (``vibes'') and methodological concepts (e.g., construct validity), developing tools that capture emergent heuristics without imposing rigid structures, and providing just-in-time methodological support. The opportunity is to meet practitioners where they are: recognizing their current practices as necessary adaptations rather than methodological failures, and providing vocabulary and structure to make their tacit knowledge explicit and shareable. This raises questions for future research. What support helps teams progress from ad-hoc to systematic evaluation without losing the signal that informal assessment provides? How can we help practitioners identify evaluation methods that fit their specific context? Are there forms of support that generalize across contexts within a given domain, for instance, does scaffolding evaluation practices in one healthcare organization transfer to others? And how can practitioners trace evaluation results to specific components they can change? We turn to this last question next.

\subsubsection{For Practice: Strategies for Confronting the Results-Actionability Gap}
The results-actionability gap, where participants obtain evaluation results but cannot translate them into action, affected 17 of 19 participants (detailed in C5). Based on patterns from successful teams in our study and examples from related work, we propose three interrelated strategies for bridging this gap:

\begin{enumerate}
    \item \textbf{Evaluation-by-design.} When evaluation is considered from the outset, it becomes an integral part of the design process rather than a post-hoc checkpoint. This does not mean rigidly adhering to initial goals, but iteratively refining both system and evaluation criteria together. Alignment activities [A8] and establishing constructs [C2] transform from late-stage struggles into ongoing design conversations. For instance, P8's team used system theoretic process analysis (STPA) not just to plan tests but to shape product requirements through evaluation thinking. This responds to \citep{zhou-etal-2022-deconstructing}'s call to ``make evaluation choices explicit'': when P6 had ``tooling in place before [they started] creating the functionality,'' evaluation insights directly influenced architecture. This integration does not just make evaluation more actionable; it also streamlines stakeholder communication (you already have shared success criteria), accelerates iteration cycles (you know what to measure after each change), and prevents costly late-stage pivots (problems surface during development, not after launch). To implement this, consider adopting P8's STPA approach, or start simpler---for each component, document: ``What's the most likely LLM failure?'' (Hallucination? Wrong tone?), ``How will we detect it?'' (User complaints? Expert review?), and ``What can we adjust if it fails?'' (Prompt? Model? Retrieval window?).
    \item \textbf{Build continuous sense-making throughout development.} Transform informal observations into institutional knowledge through lightweight documentation. Successful teams in our study maintained shared logs converting ``vibes'' into testable hypotheses (P9's ``prompt chains''), held weekly 30-minute synthesis meetings to update evaluation criteria, engaged domain experts as diagnostic partners who explain why outputs fail rather than just marking them incorrect, and documented what worked with specific changes, impacts, and decisions. We see this also in emerging industry practices where teams engage domain experts iteratively to understand failure patterns~\citep{nahar2024beyond} and maintain per-sample evaluation logs to trace what works across iterations~\citep{biderman2024lessons}. P6's team reported saving weeks by maintaining these evaluation artifacts, while teams without such practices ``reinvent[ed] the wheel'' (P15) on each project. The payoff is clear---your tenth LLM project takes a fraction of the evaluation effort of your first. Make every observation count by asking ``What would we want to know about this next time?'' and documenting the answer.
    \item \textbf{Evaluate through incremental changes.} When evaluation reveals problems, resist overhauling everything. Change one variable at a time (e.g., just the prompt or the temperature) and measure impact before the next change. Document each micro-experiment simply: change, date, impact, keep/revert. P9 noted their team often ``throw everything out and start over'' when they cannot isolate problems; incremental evaluation prevents this costly pattern. This practice, also known as \textit{satisficing} has been extensively discussed as a successful strategy for designing complex sociotechnical systems~\citep{norman2016designx, Maden_Lomas_Hekkert_2024, fokkinga2020impact}.
\end{enumerate}

These strategies address different facets of the results-actionability gap: evaluation-by-design integrates evaluation into the design process rather than treating it as an afterthought, continuous sense-making builds capacity to interpret why outputs fail, and incremental testing traces results to specific changes.

\subsection{Limitations and Future Work}
The sample size ($N=19$) aligns with typical qualitative HCI research, though recruitment through professional networks may skew toward practitioners already engaged with evaluation questions. This provides depth of insight into active practices but may not capture teams who have abandoned formal evaluation entirely. Related, our sample's gender distribution reflects typical AI/ML industry demographics but may influence findings, as evaluation approaches and risk perception can vary across demographic groups. 

Participants spanned diverse organizational contexts—from startups to Fortune 500 companies, across healthcare, legal, education, and enterprise sectors---and data collection occurred during rapid technological change (February--May 2025). While we did not systematically analyze differences between organizational contexts, the challenges we document appeared consistent; a five-person startup and a large enterprise described similar struggles with alignment, construct definition, and translating results into action. This consistency, despite organizational diversity and a fast-moving technological landscape, suggests the challenges stem from the current state of the LLM paradigm. Future work might employ longitudinal or comparative designs to examine whether these patterns persist as the technology matures and whether organizational factors systematically shape evaluation approaches.

Finally, our analysis relied on self-reported practices, capturing what participants say they do rather than observed behavior. Ethnographic methods could reveal gaps between reported and actual practices, particularly around informal activities like vibe checks. We also acknowledge that our positionality shapes interpretation (see \autoref{app:positionality}): the emphasis on social and organizational aspects, and framing certain challenges as fundamental rather than transitional, reflects our analytical lens. 

\section{Conclusion}
This study investigated how practitioners evaluate LLM-based products in production settings. Through interviews with 19 practitioners across diverse sectors, we identified ten evaluation practices and five challenges, revealing that the struggles teams face stem not from methodological failure but from structural characteristics of LLM-based systems.

Our central finding, the results-actionability gap, explains why evaluation difficulties persist even when teams successfully gather data: practitioners cannot trace poor results to specific components they can change. This gap affected 17 of 19 participants and helps account for why better metrics alone cannot solve LLM evaluation challenges. The problem is not measurement---it is that measurement often fails to provide actionable signal in systems where its components interact unpredictably.

We argue, then, that the interpretive practices we observed may be necessary adaptations to LLM characteristics rather than methodological failures. They persist not because teams await better metrics, but because they capture what metrics cannot. For HCI, this reframing suggests an opportunity: rather than developing new evaluation frameworks, the field can support practitioners in systematizing the approaches they are already developing through trial and error.

\begin{acks}
This research was partially funded by Danish Novo Nordisk Foundation under Grant Number NNF20OC0066119 and the Science of trustworthy AI award from Schmidt Sciences.
\end{acks}

%

\bibliographystyle{ACM-Reference-Format}
\bibliography{01_references}


\end{document}